\documentclass[structabstract]{aa}  
\usepackage{graphicx}
\usepackage{txfonts}
\usepackage{natbib}
\begin{document}
   \title{Interstellar HOCN in the Galactic center region\thanks{Based on observations with the IRAM 30-m telescope. IRAM is 
supported by CNRS/INSU (France), the MPG (Germany) and the IGN (Spain).}}

   \author{S. Br\"unken
          \inst{1}
          \and
          A. Belloche \inst{2}
          \and
          S. Mart\'in \inst{3,4}
          \and
          L. Verheyen \inst{2}
          \and
          K. M. Menten \inst{2}
          }

   \institute{I. Physikalisches Institut, Universit{\"a}t zu K{\"o}ln, Z{\"u}lpicher Str. 77, 50937 K{\"o}ln, Germany\\
              \email{bruenken@ph1.uni-koeln.de}
         \and
             Max-Planck-Institut f{\"u}r Radioastronomie, Auf dem H{\"u}gel 69, 53121 Bonn, Germany\\
         \and    
             Harvard-Smithsonian Center for Astrophysics, 60 Garden St., Cambridge, MA 02138, USA\\
           \and    
            European Southern Observatory, Alonso de C\'ordova 3107, Vitacura, Casilla 19001, Santiago 19, Chile\\
             }

   \date{Received ; accepted }

 
  \abstract
   {}
   {Our aim is to confirm the interstellar detection of cyanic acid, HOCN, in the Galactic center clouds. It has previously been tentatively detected only in Sgr B2(OH).}
   {We used a complete line survey of the hot cores Sgr B2(N) and (M) in the 3 mm range, complemented by additional observations carried out with the IRAM 30~m telescope at selected frequencies in the 2 mm band and towards four additional positions in the Sgr B2 cloud complex in the 2 and 3 mm bands. The spectral survey was analysed in the local thermodynamical equilibrium approximation (LTE) by modeling the emission of all identified molecules simultaneously. This allowed us to distinguish weak features of HOCN from the rich line spectrum observed in Sgr B2(N) and (M). Lines of the more stable (by 1.1~eV) isomer isocyanic acid, HNCO, in these sources, as well as those of HOCN and HNCO towards the other positions, were analysed in the LTE approximation as well.}
   {
   Four transitions of HOCN were detected in a quiescent molecular cloud in the Galactic center at a position offset in (R.A., decl.) by (20$^{\prime\prime}$,100$^{\prime\prime}$) from the hot core source Sgr B2(M), confirming its previous tentative interstellar detection. Up to four transitions were detected toward five other positions in the Sgr B2 complex, including the hot cores Sgr B2(M), (S), and (N). A fairly constant abundance ratio of $\sim 0.3-0.8$~\% for HOCN relative to HNCO was derived for the extended gas components, suggesting a common formation process of these isomers. }
   {}


   \maketitle
%

\section{Introduction}
Ever since its first detection towards Sgr B2 (OH) \citep{SB1972}, isocyanic acid (HNCO) has been found with high abundances in a multitude of galactic sources, such as high-mass star-forming regions \citep{JAB1984,ZHM2000,BJD2007}, Galactic center molecular clouds \citep{DHW1997,MRM2008}, as well as translucent and dark clouds \citep {TTH1999}. HNCO has also been detected in several external galaxies \citep{NHJ1991,WHC2004,MMM2009}, and been suggested as a tracer for nuclear starburst activity \citep{MRM2008,MMM2009}. Two energetically less stable isomers of HNCO have recently been detected in the interstellar medium: cyanic acid (HOCN) tentatively towards Sgr B2 (OH) following its laboratory identification \citep{BGM2009b}, and fulminic acid (HCNO) towards three starless dark clouds and one low-mass star-forming region \citep{MCT2009} with abundance ratios relative to HNCO of 0.5\% and $1-5$\%, respectively. The relative abundances of members of an isomeric family reflect their different formation and destruction processes, and observations of isomers of  small systems like HCN/HNC  \citep[e.g., ][]{SWP1992} as well as complex organic species \citep{HLJ2000,TA2001} have been previously used to investigate their gas-phase and/or grain-surface interstellar origin. 

The Giant Molecular Cloud Sgr B2 at a projected distance of $\sim$90~pc from the Galactic center, where HOCN had been tentatively detected, harbors three massive star-forming regions associated with the most massive (sub)mm continuum sources in this region: Sgr B2(N), (M), and (S). These hot cores are aligned in a north-south direction in the dense central $5-10$ pc of Sgr B2 and have been shown to be extremely rich in organic molecules by extensive molecular line surveys \citep[e.g., ][]{CLT1986,Turner1991,NBH2000,BMC2008a,BMC2008b,BGM2009a}.  Whereas most molecules have peak abundances in the hot cores, HNCO emission was found to be extended with an expanding ringlike morphology and a peak at a position 2' north (2'N) of Sgr B2(M), where only a few other molecules, like HCO$_2^+$ and CH$_3$OH, have enhanced abundances \citep{MHH1998,MI2006,JBC2008}. The HNCO enhancement at this particular position has been explained by a cloud-cloud collision, resulting in shocks that release grain mantle material which is further processed via gas-phase reactions \citep{MI2006}. Similarly, large-scale, low-velocity shocks are proposed as responsible for the large abundances of complex organic molecules, including HNCO, in the quiescent molecular clouds throughout the whole Galactic-center region \citep{RMR2006,RMM2008}. 

In this paper we present additional astronomical observations of HOCN toward several positions in the Sgr B2 complex with the IRAM 30m telescope. These observations confirm the tentative identification by \cite{BGM2009b} based on previously published line surveys of Sgr B2(OH). We find evidence of HOCN towards six positions, including the hot cores Sgr B2(M) and (N). 
The interstellar detection of HOCN is unequivocally confirmed by the observation of four unblended transitions toward a position very close to the 2'N cloud, at an offset in (R.A., decl.) by (20$^{\prime\prime}$,100$^{\prime\prime}$) from Sgr B2(M). We compare the relative abundances of HOCN to the most stable isomer HNCO.


\section{Astronomical observations and analysis}

\begin{figure*}
\includegraphics[width=8.0cm]{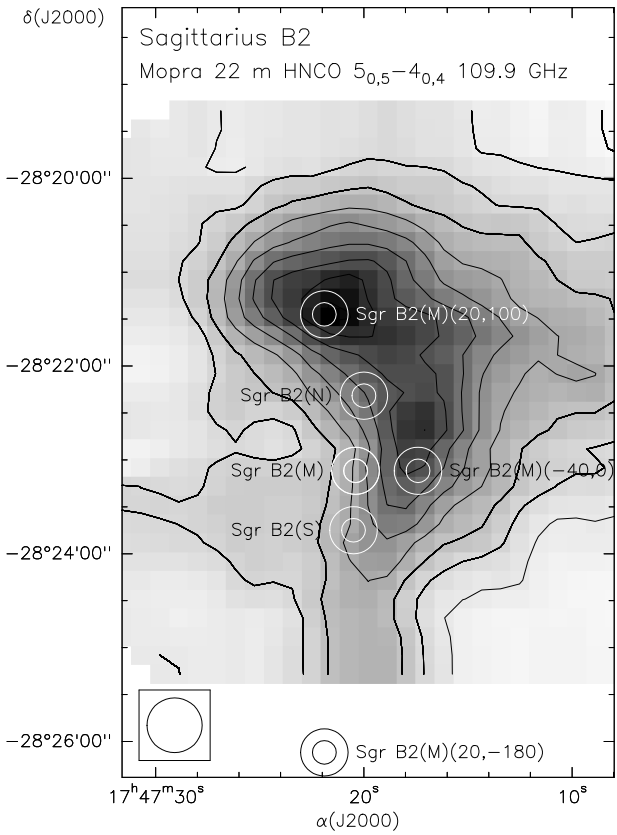}
\caption{The gray-scale and contours showing the velocity-integrated emission from the 109.9 GHz 
$5_{0,5}-4_{0,4}$ transition of HNCO in the Sgr B2 region imaged by Jones et
al. (2008) with the 22 meter Mopra telescope. Contours start at and are in steps of
10~Kkm~s$^{-1}$ in the Mopra $T_{\rm A}^*$ scale. The Mopra telescope has an FWHM beam width of $35''$ at 109.9
GHz, which is represented at the lower left corner.
The other circles are centered on the positions toward which we took HOCN
spectra with the IRAM 30 m telescope (see Tables 1 and 2 and Figs. 2--5). For
each position, the diameter of the large and the small circles represent the
FWHM beam widths at the lowest and the highest frequencies of an HOCN line
observed by us, 83.9 and 167.8 GHz, respectively, i.e., $30''$ and $15''$. Note
how the HNCO emission ``avoids'' the Sgr B2(M) and (N) dense, hot molecular
cores.}
\label{fig_positions}
\end{figure*}

All astronomical observations were done with the IRAM 30~m telescope on Pico Veleta (Spain), but during different observing campaigns.
We initially searched for astronomical lines of HOCN in existing line survey data in the 3~mm band between 80 and 116 GHz toward the hot core regions Sgr B2(N) and (M). The observations were done in January and September 2004, and January 2005. Details of the observation and data reduction of this sensitive survey have been given by \cite{BMC2008a,BMC2008b}. Additional spectra were obtained in May 2008 in the 2 and 3 mm bands toward Sgr B2(N) and Sgr B2(M), and a position (20$^{\prime\prime}$,100$^{\prime\prime}$) offset from Sgr B2(M), a molecular cloud region in the vicinity (at an angular offset of 28$^{\prime\prime}$) of the HNCO peak position Sgr B2(2$^{\prime}$N) \citep{MI2006}. The spectral resolution of these observations was 0.313~MHz. We also include data from a survey toward several additional positions in the Galactic center region observed with the 30~m telescope in July 2004 and May 2005, with a lower spectral resolution of 1~MHz and 4~MHz at 3~mm and 2~mm, respectively (see \cite{MRM2008} for details). The observed positions are shown in Fig. \ref{fig_positions} in comparison to the large-scale HNCO emission as measured by \cite{JBC2008}, and the coordinates are given in Table \ref{table_positions}.

A common problem encountered in the detection of new molecules in the Sgr B2 region is to distinguish between lines of the target molecule and those of other species. 
Since the two hot-core regions Sgr B2(N) and (M) are both extremely rich molecular sources, with about 100 lines and 25 lines per GHz over the 3~mm band, respectively, it is crucial to model the emission of all identified molecules in these sources simultaneously to avoid misassignments and to identify possible line overlaps. 
For this purpose our complete 3~mm survey data of these two sources was analyzed using the XCLASS software developed by P.~Schilke\footnote{http://www.astro.uni-koeln.de/projects/schilke/XCLASS}, which allows the simultaneous fit of multiple transitions and molecules under the assumption of local thermodynamic equilibrium (LTE). With this method, we generated an LTE model of the emission of all hitherto identified molecules, as well as of HNCO and HOCN. The free parameters in this model are source size, column density, rotational temperature, velocity offset and line width, with a specific set of parameters for each molecule. This is necessary since the kinetic temperature and the molecular distribution in the Sgr B2 sources are not uniform: large variations in the rotational temperatures are commonly found for different molecules in these sources \citep[e.g.,][]{NBH2000}. Furthermore, for the same molecule, multiple velocity and temperature components have been found in several cases \citep[e.g.,][]{HPB2003,HJL2004,BGM2009a}. This was taken into account in the analysis by modeling the molecular emission where the line profile departed significantly from a single Gaussian with several Gaussian velocity components, i.e., several distinct parameter sets. 

Details of this procedure for the Sgr B2(N) region have been given previously 
\citep{BMC2008a,BGM2009a,MBM2008}, whereas the model of Sgr B2(M) will be described elsewhere (A. Belloche {\it et al.} in preparation). 
Our LTE modeling assumes a very basic source structure, and the radiative 
transfer calculations are approximate. Each emission/absorption component can 
have a specific size and has uniform physical parameters (temperature, column 
density, linewidth, velocity). The contributions from different components are 
linearly added together, which is a correct assumption for, e.g., two 
non-overlapping compact sources smaller than the beam, but is not a good 
approximation for, e.g., optically thick lines in the case of two sources that 
overlap along the line of sight. Finally, the absorption components are marked 
with a specific flag and calculated using the continuum emission 
\textit{and} the emission lines as background. 
Our models of the hot cores Sgr~B2(N) and (M) include a combination of a warm 
and a cold component that most likely overlap along the line of sight. If one 
line of the cold component is optically thick, we overestimate its intensity 
by simply adding the contribution from the warm component that lies behind. In 
the following, this is not an issue for HOCN because all components in our LTE 
model are optically thin. But it is somewhat inaccurate for HNCO because the 
foreground cold component of our model for Sgr~B2(N) has opacities up to 1 in 
the 3 mm atmospheric window.

The XCLASS software computes the brightness temperature of the LTE model of 
one component with the formula
\begin{equation}
T_{\nu} = \eta_{\nu} \times (J_{\nu}(T_{\mathrm{ex}}) - J_{\nu}(T_{\mathrm{bg}})) \times (1 - exp(-\tau_{\nu})),
\end{equation}
with $\eta_{\nu}$ the beam filling factor, $T_{\mathrm{ex}}$ the excitation 
temperature (equal to the kinetic temperature since we assume LTE conditions), 
$T_{\mathrm{bg}}$ the background temperature, and $\tau_{\nu}$ the opacity of 
the line at frequency $\nu$. Here we used $T_{\mathrm{bg}} = 2.7$ K, i.e. the 
cosmic microwave background, for all sources except Sgr B2(N) and (M). 
For modeling HOCN and HNCO toward Sgr~B2(N) and (M), we used the same 
background temperatures as for the analysis of the whole survey \citep{BMC2008a,BGM2009a}: 
$T_{\mathrm{bg}} =$ 5.2 and 5.9~K at 3~mm, 6.5 and 7.0~K at 
2~mm, 10.0 and 8.5 K at 1~mm, respectively. These background temperatures were 
derived from the saturated \textit{absorption} lines produced by the 
foreground diffuse clouds assuming a beam filling factor of 1. They were then 
in turn also used for all \textit{emission} lines. This should be a reasonable 
assumption for the extended \textit{emission} components if the continuum 
source has a size comparable to or larger than the beam. It may not be very 
accurate for the compact \textit{emission} components that likely see a 
stronger background. However, the continuum emission is still optically thin
at 3 mm, so the difference $J_{\nu}(T_{\mathrm{ex}}) - J_{\nu}(T_{\mathrm{bg}})$
is not very sensitive to the exact value of $J_{\nu}(T_{\mathrm{bg}})$ as long as
it is much smaller than $J_{\nu}(T_{\mathrm{ex}})$ of the hot, compact 
components. With the current state of our software analysis, improvement of 
the background modeling (especially at 1~mm) is not straightforward and is 
beyond the scope of this article.

Previous observations have shown that there is much less line confusion at the other positions listed in Table  \ref{table_positions} , with the exception of the hot core region Sgr B2(S). We used the same method as for Sgr B2(N) and (M) to model the emission of HOCN and HNCO in the LTE approximation toward these other positions. A background temperature of 2.7~K was taken into account in these cases. The LTE models of Sgr B2 (N) and (M) can be used as a guide to resolve questions concerning possible blending of lines for these sources.

\section{Results}

\begin{figure*}
\includegraphics[width=4.0cm]{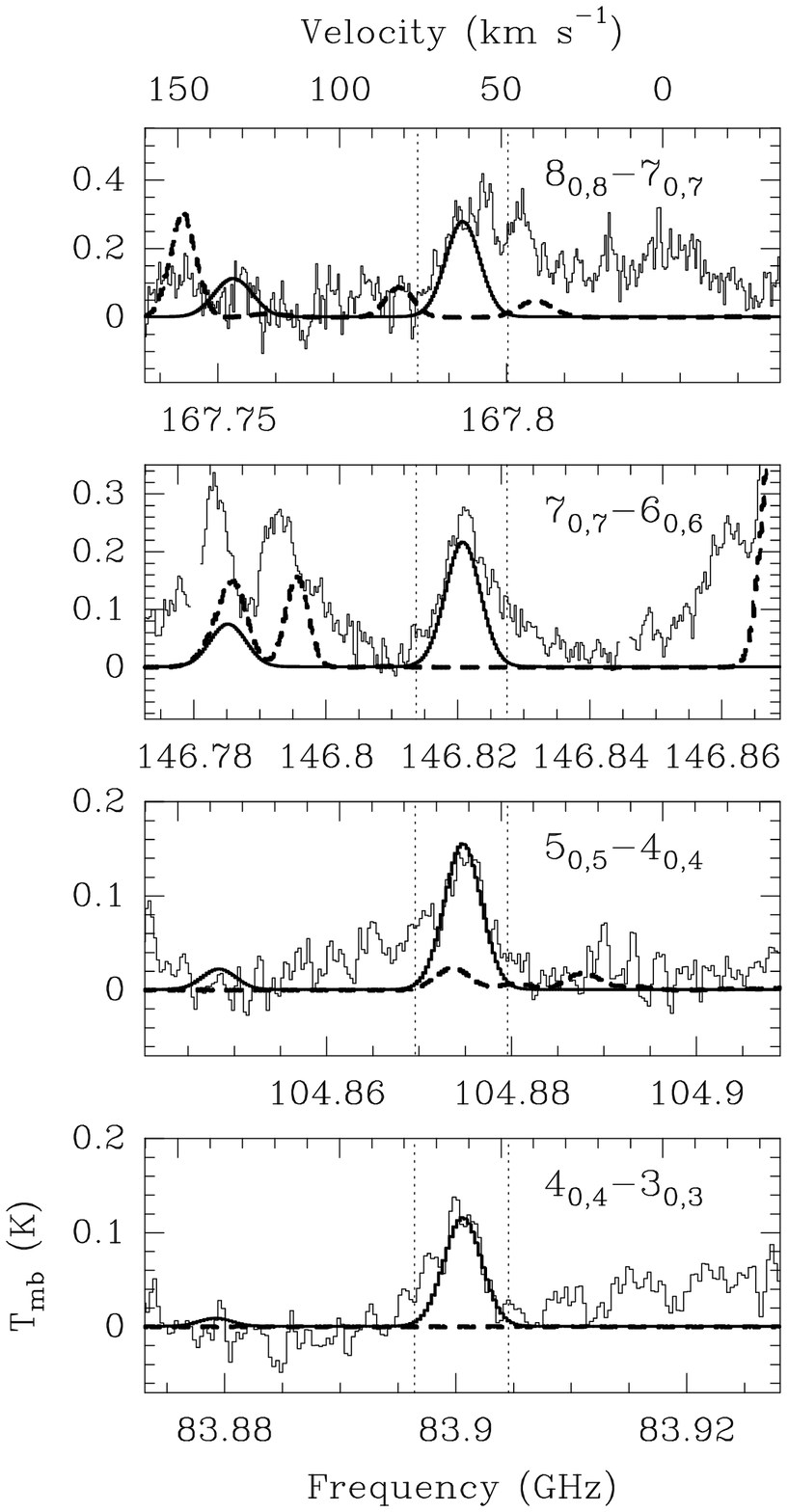}
\includegraphics[width=8.0cm]{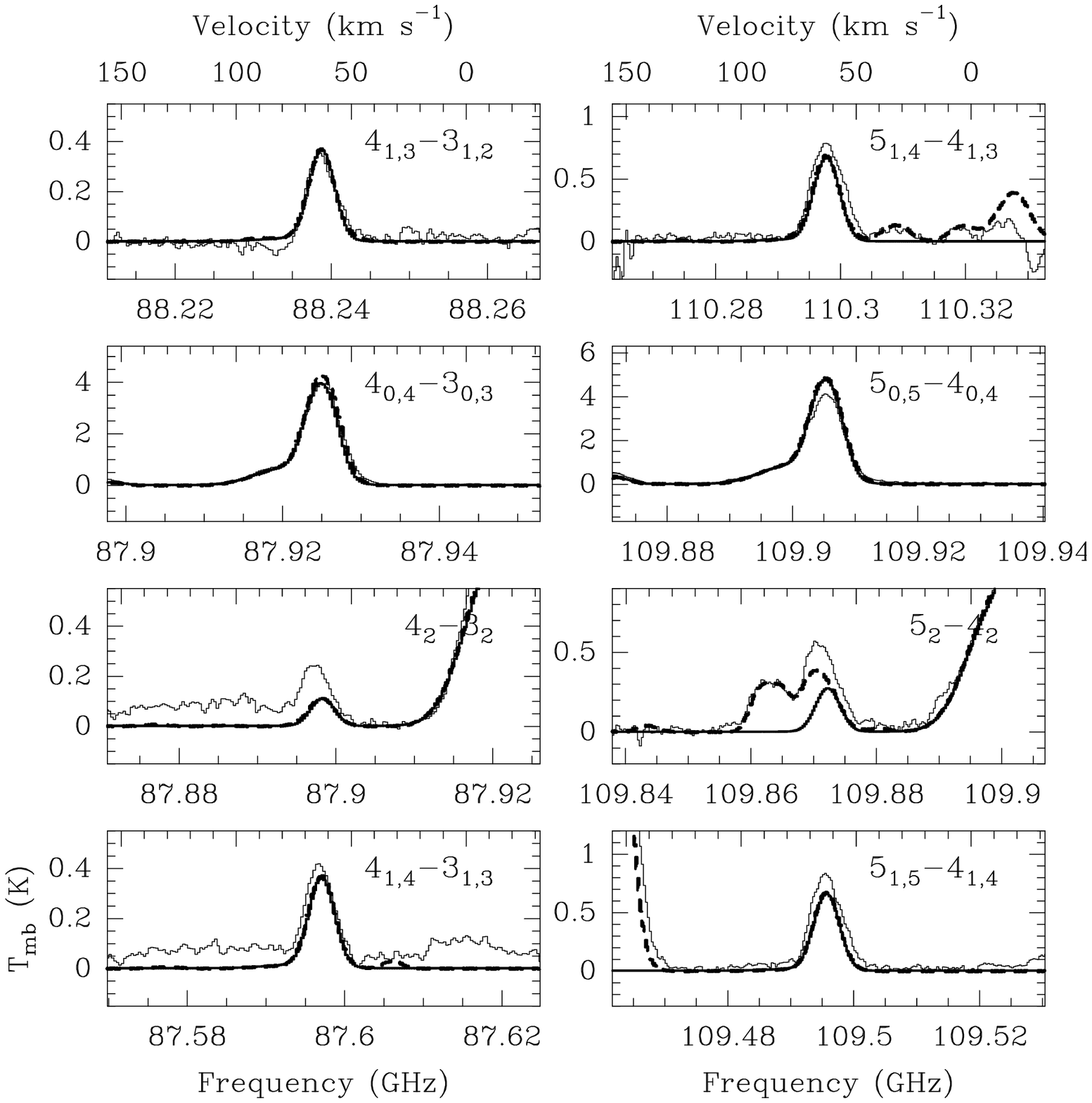}\\
\hspace{3cm} a) HOCN \hspace{3cm} b) HNCO
\caption{Detected lines of HOCN and HNCO toward Sgr B2(M). In each panel, the observed spectrum (thin line) is shown with the LTE synthetic spectrum of HOCN or HNCO, respectively (thick line), and the LTE model including all identified molecules in this source except HOCN (dashed line). The small feature in the synthetic HOCN spectrum towards lower frequencies is one of the $K_a$=2 transitions, which is only barely detected. The parameters of the LTE model are given in Table \ref{table_astropar}.}
\label{fig_M}
\end{figure*}

\begin{figure*}
\includegraphics[width=4.0cm]{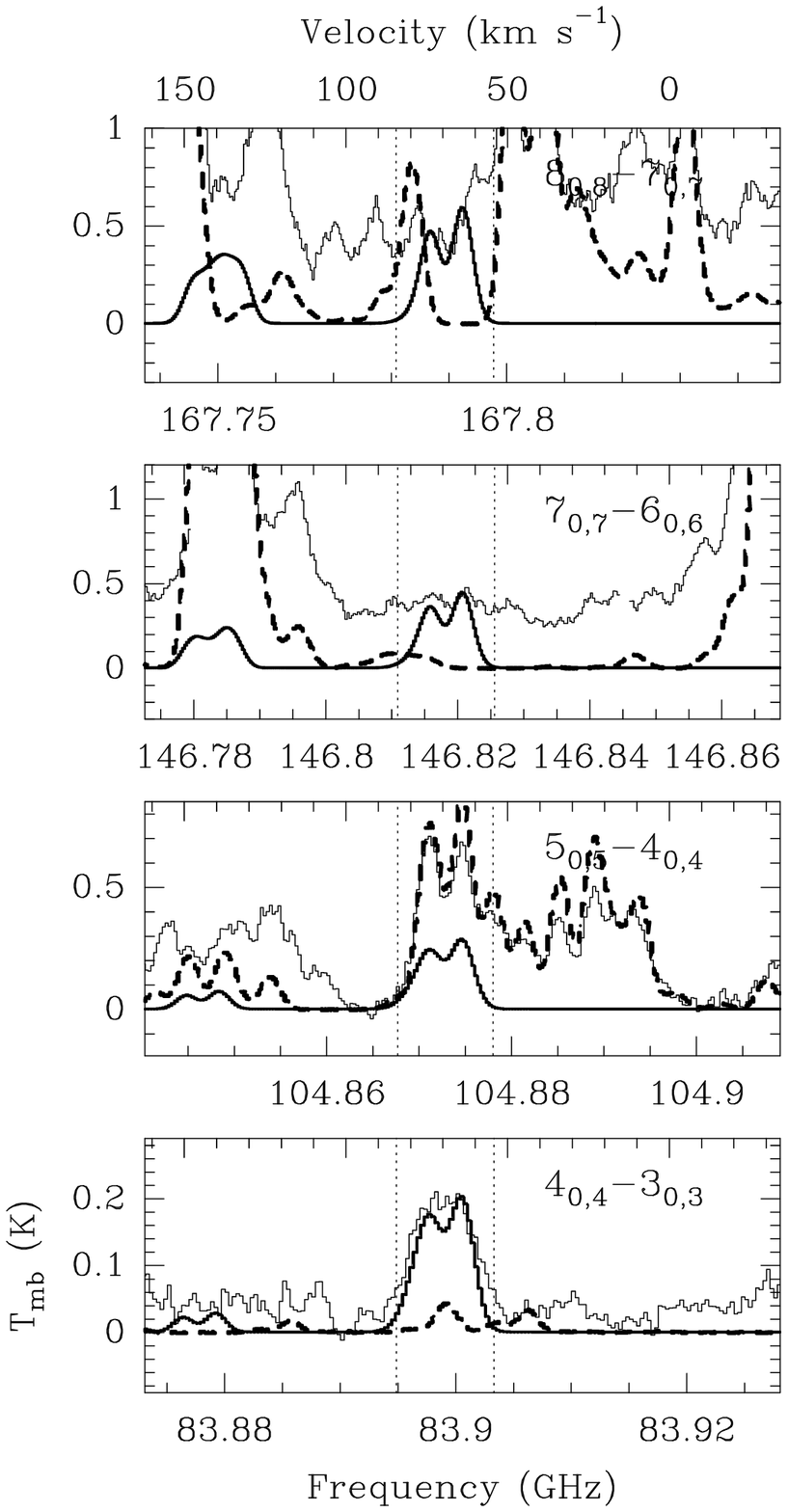}
\includegraphics[width=8.0cm]{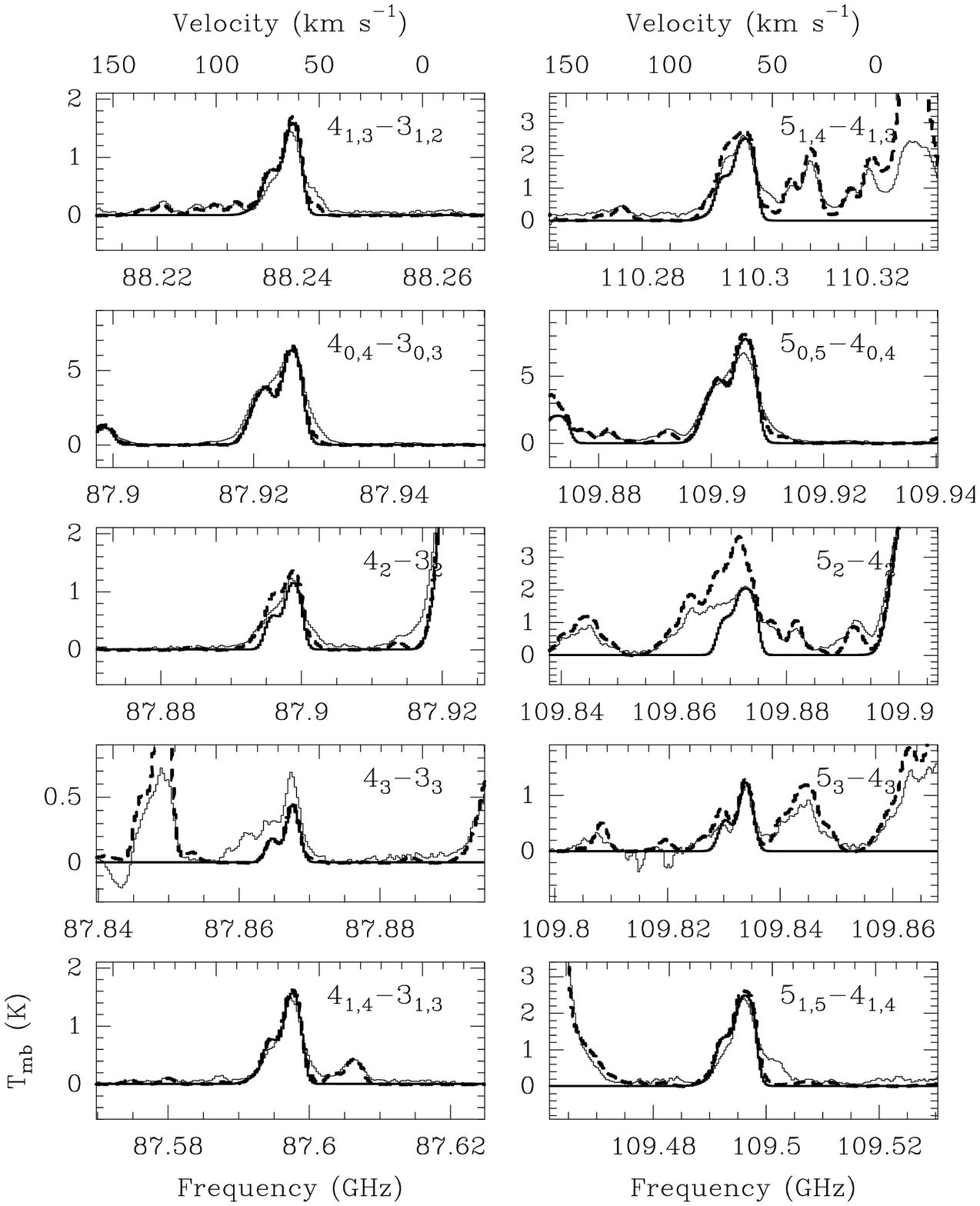}\\
\hspace{3cm} a) HOCN \hspace{3cm} b) HNCO
\caption{Detected lines of HOCN and HNCO toward Sgr B2(N). In each panel, the observed spectrum (thin line) is shown with the LTE synthetic spectrum of HOCN and HNCO, respectively (thick line), and the LTE model including all identified molecules in this source except HOCN (dashed line). The HOCN emission was modeled assuming two velocity components. The parameters of the LTE model for HOCN and HNCO are given in Table \ref{table_astropar}.}
\label{fig_N}
\end{figure*}

\begin{figure*}
\includegraphics[width=4.5cm]{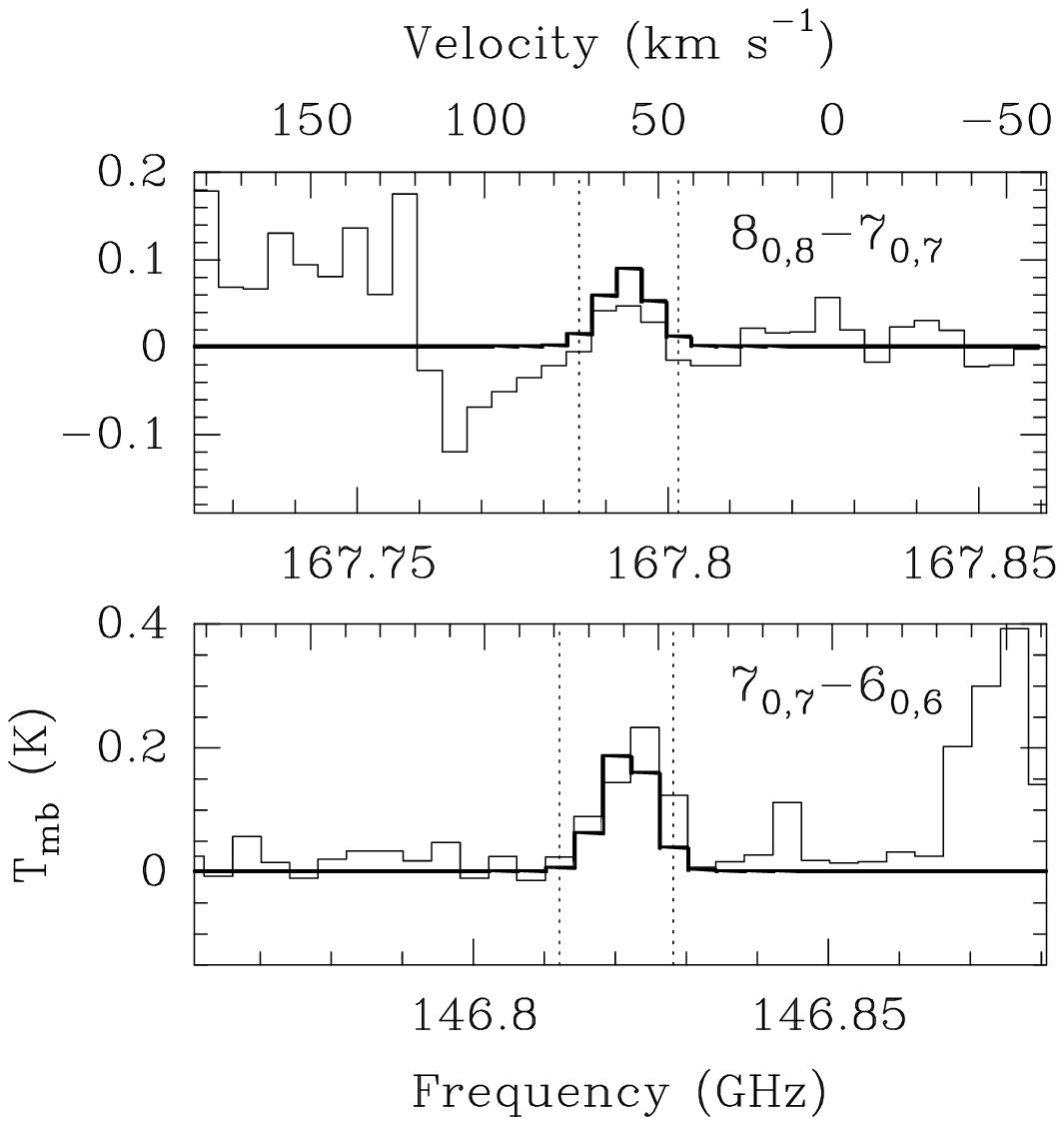}
\includegraphics[width=4.5cm]{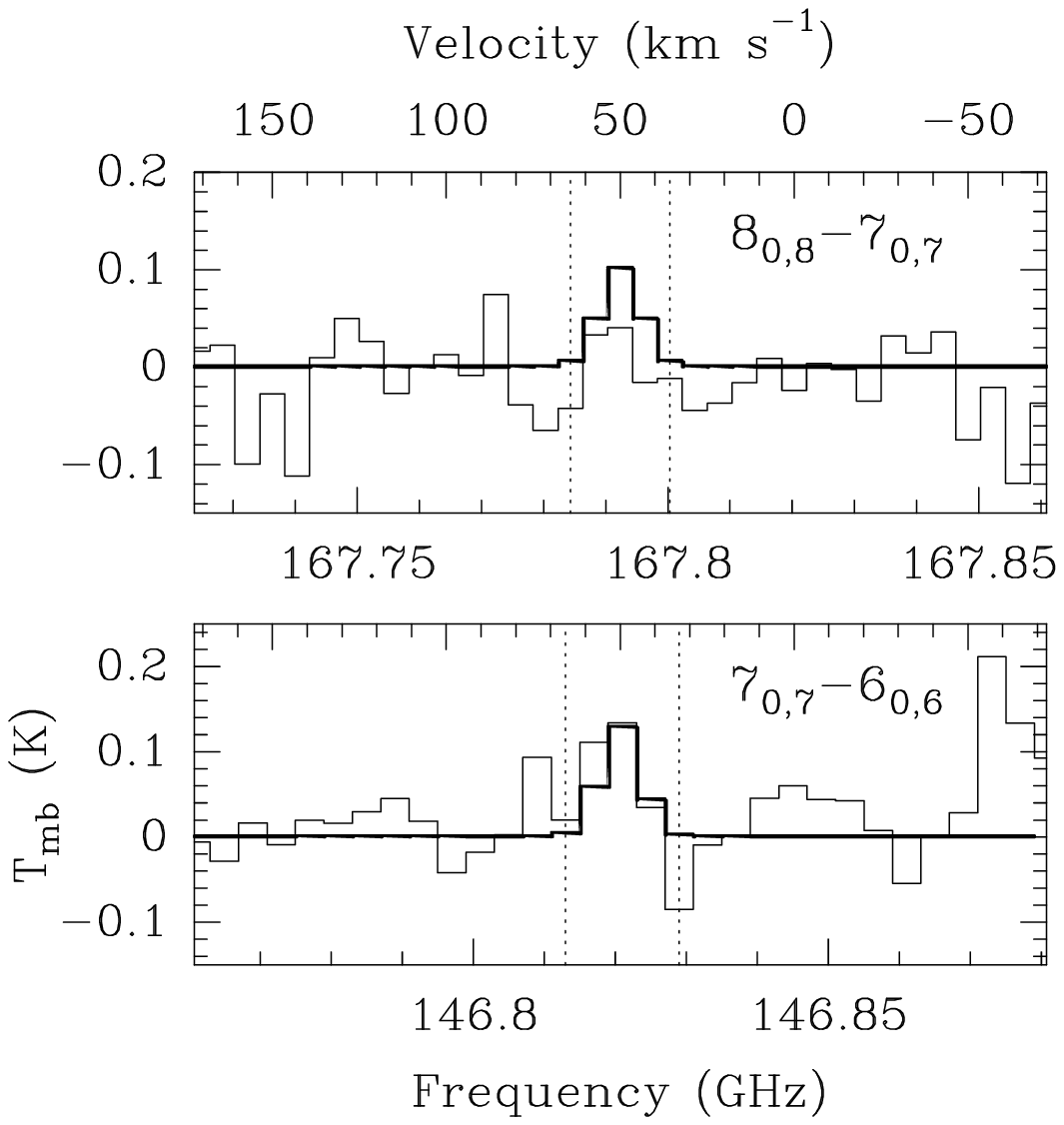}
\includegraphics[width=4.5cm]{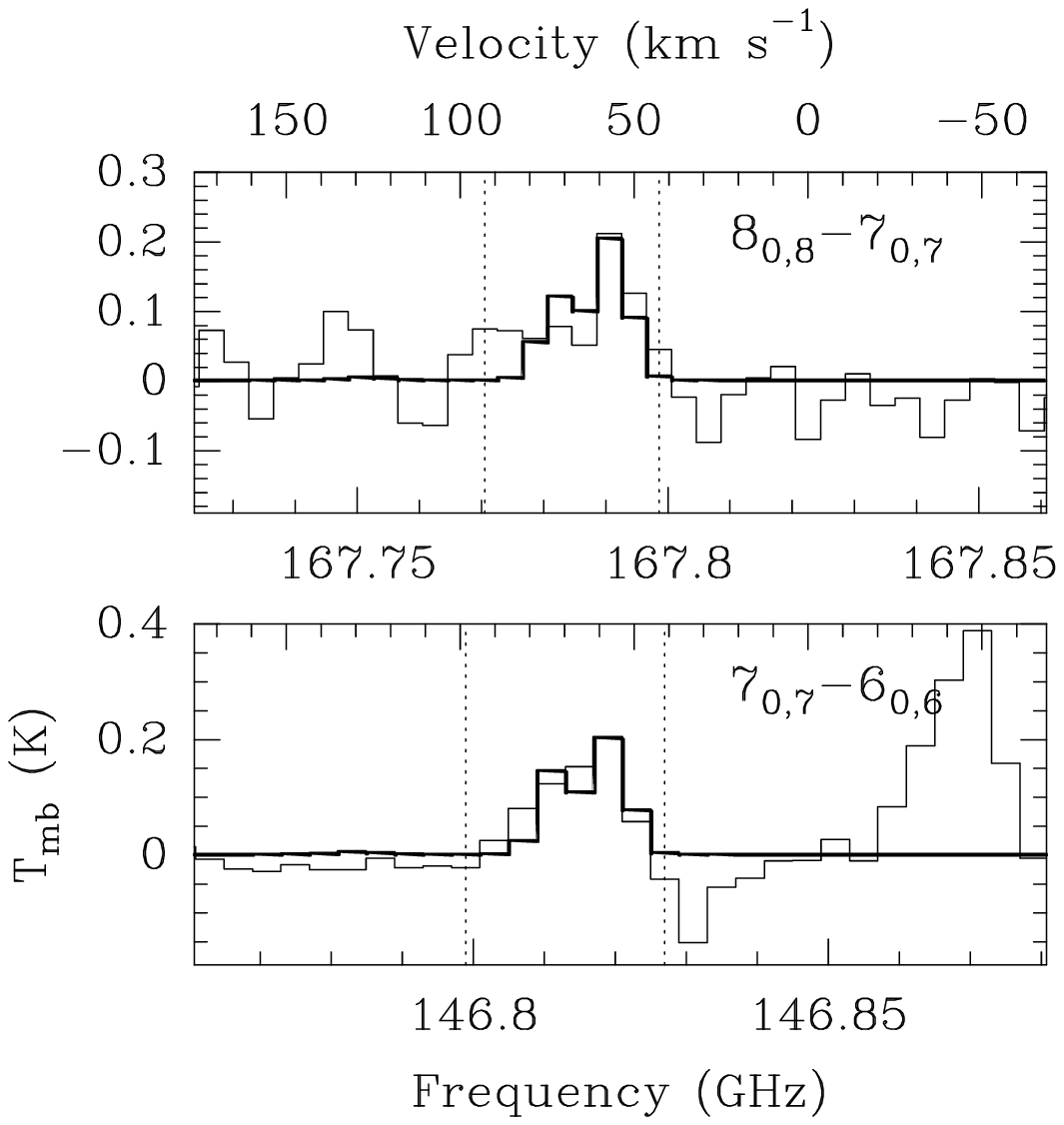}
\newline
 a) Sgr B2(S) \hspace{3.2cm} b)  Sgr B2(M)(20$^{\prime\prime}$,-180$^{\prime\prime}$) \hspace{1.2cm} c) Sgr B2(M)(-40$^{\prime\prime}$,0$^{\prime\prime}$) 
\caption{Observed lines (thin line) of HOCN toward Sgr B2(S),  (M)(20$^{\prime\prime}$,-180$^{\prime\prime}$), and (M)(-40$^{\prime\prime}$,0$^{\prime\prime}$) with the overlaid LTE synthetic spectrum of HOCN (thick line) based on the model given in Table \ref{table_astropar}.}
\label{fig_offsetpositions}
\end{figure*}

The results of our search for HOCN toward Sgr B2 are shown in Figs. \ref{fig_M20,100}, \ref{fig_M}, and \ref{fig_N}, \ref{fig_offsetpositions} and are summarized in detail in Table \ref{table_detecthocn}. Four $a$-type transitions of HOCN in the $K_a=0$ ladder were detected with consistent line intensities and widths at the predicted frequencies toward Sgr B2(M) and (M)(20$^{\prime\prime}$,100$^{\prime\prime}$). In the much more line-rich and, thus, confusion prone hot-core region Sgr B2(N), only the $4_{0,4}-3_{0,3}$ transition has been found to be free of contaminating emission from other species. The $7_{0,7}-6_{0,6}$ and $8_{0,8}-7_{0,7}$ transitions of HOCN were covered in the survey toward the remaining positions Sgr B2(S), (M)(-40$^{\prime\prime}$,0$^{\prime\prime}$), and (M)(20$^{\prime\prime}$,-180$^{\prime\prime}$). Whereas the lower frequency line was clearly detected toward the first two positions, and at about the 3$\sigma$-level toward the third, the detection of the higher frequency transition is more uncertain because the high noise level of these observations. However, the detection is plausible in these positions since no contamination by other species is expected at these frequencies, as observed in Sgr B2(M)(20$^{\prime\prime}$,100$^{\prime\prime}$), and the LTE predictions are consistent with the observations.  

Our best-fit LTE model parameters for HOCN and HNCO are summarized in Table \ref{table_astropar}. The source size was fixed at 60$^{\prime\prime}$ for the extended emission, based on the observed large-scale distribution of HNCO \citep{CWM1986,MI2006}. Source sizes of 2$^{\prime\prime}$--3$^{\prime\prime}$ were determined from the HNCO observations for the hot compact components in the two hot-core sources Sgr B2(M) and (N). The fitting of the model to the observed spectra was done by eye using XCLASS, 3$\sigma$ uncertainties for the column densities and rotational temperatures were estimated whenever possible by varying these model parameters for each molecule and component until the agreement with the measured spectra worsened. Whereas this method gives considerably larger column density uncertainties than the statistical errors obtained from the standard method using rotational Boltzmann diagrams, these values are more realistic since they also account for systematic uncertainties owing to apparent line blendings and the existence of multiple source components.

In the following we give some details of the modeling for the different sources.

\noindent
{\it Sgr B2(M)(20$^{\prime\prime}$,100$^{\prime\prime}$)}:  \newline
The line profiles of HOCN and HNCO toward Sgr B2(M)(20$^{\prime\prime}$,100$^{\prime\prime}$) are complex, which is accounted for by assuming three velocity components ($v_{LSR}=63$, 71, and 79~km/s) in the analysis. For comparison the observed \citep{MRM2008} and synthetic model spectra of HNCO are also shown in Fig. \ref{fig_M20,100} for the same velocity components as for HOCN. The overall lineshape of HNCO emission is reproduced well, if the lower spectral resolution of the observations is taken into account. There is an additional blueshifted wing component in the HNCO line profiles, but we did not attempt to model it with an additional velocity component since it is barely detected in only one HOCN transition. \newline
{\it Sgr B2(M)}: \newline
A two-component model -- a compact hot core ($T_{rot}=100$~K) and an extended cold envelope ($T_{rot}=12$~K)-- with one velocity component at 62~km/s was used for Sgr B2(M).  The temperature of the hot component for HNCO was fixed to what was obtained for HOCN. \newline
{\it Sgr B2(N)}: \newline
Only one unblended line of HOCN was detected toward Sgr B2(N), so that the model parameters (except the column density) were fixed to those obtained for HNCO. Two distinct velocity components were detected at 64~km/s and 75~km/s, which had been previously observed for other molecules in this source \citep[e.g.,][]{HPB2003,BMC2008a,MBM2008,BGM2009a}. A much weaker red wing at 82~km/s was seen for HNCO, but it is barely detected for HOCN so it was not included in the analysis. For each of the velocity components, a compact hot core ($T_{rot}=200$~K) and an extended cold envelope with $T_{rot}=14$~K, obtained from the HNCO observations, were assumed. \newline
{\it Sgr B2(M)(20$^{\prime\prime}$,-180$^{\prime\prime}$) and Sgr B2(S)}: \newline
A simple one-component, one-velocity model of HOCN emission was used for Sgr B2(M)(20$^{\prime\prime}$,-180$^{\prime\prime}$) and Sgr B2(S), owing to the limited data. The rotational temperature was fixed to that of HNCO for Sgr B2(M)(20$^{\prime\prime}$,-180$^{\prime\prime}$) \citep{MRM2008}, but a much lower HOCN temperature with respect to HNCO was needed for Sgr B2(S) to reproduce the observed spectra. HNCO column densities were taken from \cite{MRM2008} for these sources, after correcting them for beam efficiencies and beam filling factors. \newline
{\it Sgr B2(M)(-40$^{\prime\prime}$,0$^{\prime\prime}$)}: \newline
The lines of both HNCO (see Fig. \ref{fig_HNCO_M-40,0}) and HOCN (Fig. \ref{fig_offsetpositions}) toward Sgr B2(M)(-40$^{\prime\prime}$,0$^{\prime\prime}$) show velocity structure, which was accounted for by a two-component model with $v_{LSR}=56.5$, and 72.2~km/s. The same velocity offsets, linewidths, and rotational temperatures were assumed for HNCO and HOCN, based upon the analysis of the former. The redshifted wing seen in the lower frequency HNCO line profiles was not detected in HOCN, so it was not modeled with an additional velocity component. 

The observed relative velocities, line widths, and rotational temperatures of HOCN generally agree with those found for HNCO, suggesting a close correlation between these two molecules. The only significant difference is found in the lower rotational temperatures of HOCN with respect to HNCO toward Sgr B2(M)(20$^{\prime\prime}$,100$^{\prime\prime}$) and Sgr B2(S), possibly because of the larger dipole moment of HOCN [$\mu _a=3.7$~D \citep{ML2008}, as compared to 1.6~D for HNCO  \citep{HGW1975}]. Einstein $A$-coefficients for $a$-type transitions are a factor five larger in HOCN than in HNCO, increasing the H$_2$ density necessary for thermalization of the rotational levels within the $K_a=0$ ladder via collisional excitation to $>2.5 \times 10^6$~cm$^{-3}$ \citep{CWM1986}, well above the typical H$_2$ densities of $\sim 10^4-10^5$~cm$^{-3}$ in the Galactic center clouds \citep{RMR2006}. We find abundance ratios of HOCN relative to HNCO of around 0.25-2.5~\% toward those sources where cyanic acid was detected (see last column of Table \ref{table_astropar}). The largest variations in this ratio are related to the estimates from the hot core components of the LTE analysis. Neglecting these components results in fairly uniform abundance ratios of 0.3-0.8~\% for the low excitation, extended molecular gas components; i.e., they agree within a factor of $\sim3$, which is a similar close agreement as found for the abundance ratios of other complex organic molecules in the Galactic center \citep{RMM2008}.

\section{Discussion and conclusion}

HOCN has previously been tentatively detected \citep{BGM2009b} based on published spectral line survey data of Sgr B2(OH) obtained with the BTL 7~m and NRAO 11~m telescopes \citep{CLT1986,Tur1989}. The current observations confirm and improve the detection of HOCN in several important ways: i) Four unblended lines of HOCN were detected in the extended gas at an offset position (20$^{\prime\prime}$,100$^{\prime\prime}$) from the hot core Sgr B2(M) close to the maximum of HNCO emission. This region does not suffer from the severe line confusion affecting the hot core regions, making incidental blendings and false assignments less likely. ii) The comparison of the observed spectrum with a synthetic spectrum -- based on the simultaneous analysis of our complete 3~mm survey data -- allows disentangling contributions of known, mostly complex, molecular species from the weak features associated with HOCN in the hot core regions Sgr B2(N) and (M). iii) The beamsize of the IRAM 30~m telescope is 2.7-times smaller than that of the NRAO 11~m telescope used by \cite{Tur1989}. This higher spatial resolution allows separation of different components in a complex source like Sgr B2. iv) The observed line intensity ratios are consistent with those predicted in the LTE approximation. v) It should also be noted that all lines of HOCN with an expected intensity over the noise level have either been detected or are blended  in the covered frequency range towards the six selected sources; i. e., no expected lines are missing.

HNCO is ubiquitous throughout the Central Molecular Zone (CMZ) of our Galaxy \citep{Menten2004}, stretching from longitudes of $-0.2$ to 1.8 degrees over a $\sim$300~pc long band along the Galactic plane \citep{DHW1997}. In Sgr B2 -- the most prominent molecular emission region within the CMZ -- the molecule exhibits large-scale abundance variations peaking not at the hot core positions -- where most other molecules show strong emission -- but rather has an extended ring-like structure, investigated in detail by \cite{MI2006} and \cite{JBC2008}. The enhancement may result from shock evaporation of icy grain mantles caused by the collision between the principal cloud and the ring, which appears to be expanding. Detection of cyanic acid toward several chemically and physically distinct regions in Sgr B2 with a fairly constant abundance ratio relative to the most stable isomer indicates an effective formation process closely related to that of HNCO. On the basis of this limited sample, HOCN may be as widespread as HNCO. 

Current astrochemical gas-grain model calculations efficiently produce HNCO on grain surfaces by hydrogenation of accreted OCN \citep{HH1993,GWH2008}. The observed fairly constant abundance ratio of HOCN relative to HNCO  in our sample of sources, which is far from thermodynamical equilibrium considering the low temperatures prevalent in the interstellar medium\footnote{HOCN has been calculated to be around 1.1~eV higher in energy than HNCO \citep{ML2008}, even at 200~K this results in an equilibrium ratio for HNCO/HOCN of $6\times 10^{27}$.}, suggests that HOCN is formed along with HNCO on the grain mantles, and then subsequently released through shocks or during warm-up. 
Motivated by the recent detections of HCNO \citep{MCT2009} and HOCN \citep[][and this work]{BGM2009b}, E. Herbst and coworkers are currently extending the Ohio State gas-grain network to account for the formation and destruction of HNCO and its higher lying isomers in different types of sources (Quan {\it et al.}, in preparation). 
Their hot core and warm envelope models can account reasonably well, i.e. with an order-of-magnitude agreement,  for the relative and absolute abundances of HNCO and HOCN observed here in the respective Sgr B2 components.  Discrepancies might come from shock chemistry (see above), which has so far not been implemented in the model calculations. From the present observations, it cannot be conclusively decided whether grain-surface chemistry plays a dominant role in the synthesis of CHNO isomers. Pure gas-phase chemical models, based on the steady-state approach used by \cite{MCT2009}, and extended by reactions involving protonated forms of NCO, CNO, and of the four stable CHNO isomers, predict HNCO to HOCN abundance ratios matching the observed values stated in Table \ref{table_astropar} for the Galactic center clouds \citep{MBC2010}

The abundances of HNCO and its isomers are governed by a complex network of surface and gas-phase formation and destruction reactions. 
Further observations of HOCN toward additional regions of the interstellar medium rich in HNCO such as other high-mass star-forming regions, young stellar objects (YSOs), and dense cloud cores -- and the observation of HOCN in sources where HCNO has been detected and vice versa -- will help to further constrain the interstellar chemistry of this isomeric system. These studies might also be extended to related isomeric families.
The higher energy isomers of the insterstellar molecule HNCS, which is isoelectronic to HNCO, have recently been detected in the laboratory \citep[HSCN;][]{BYG2010}, (HCNS and HSNC; M. C. McCarthy, in preparation), and HSCN has now also been detected in Sgr B2 \citep{HZB2009} . 

\begin{acknowledgements}
  We wish to thank E.~Herbst, R.~Garrod, and D.~Quan for helpful discussions regarding the involved astrochemistry. We thank the referee for providing constructive comments.    
\end{acknowledgements}

\bibliographystyle{aa}

\clearpage

\clearpage

\begin{figure*}
\includegraphics[width=7.6cm]{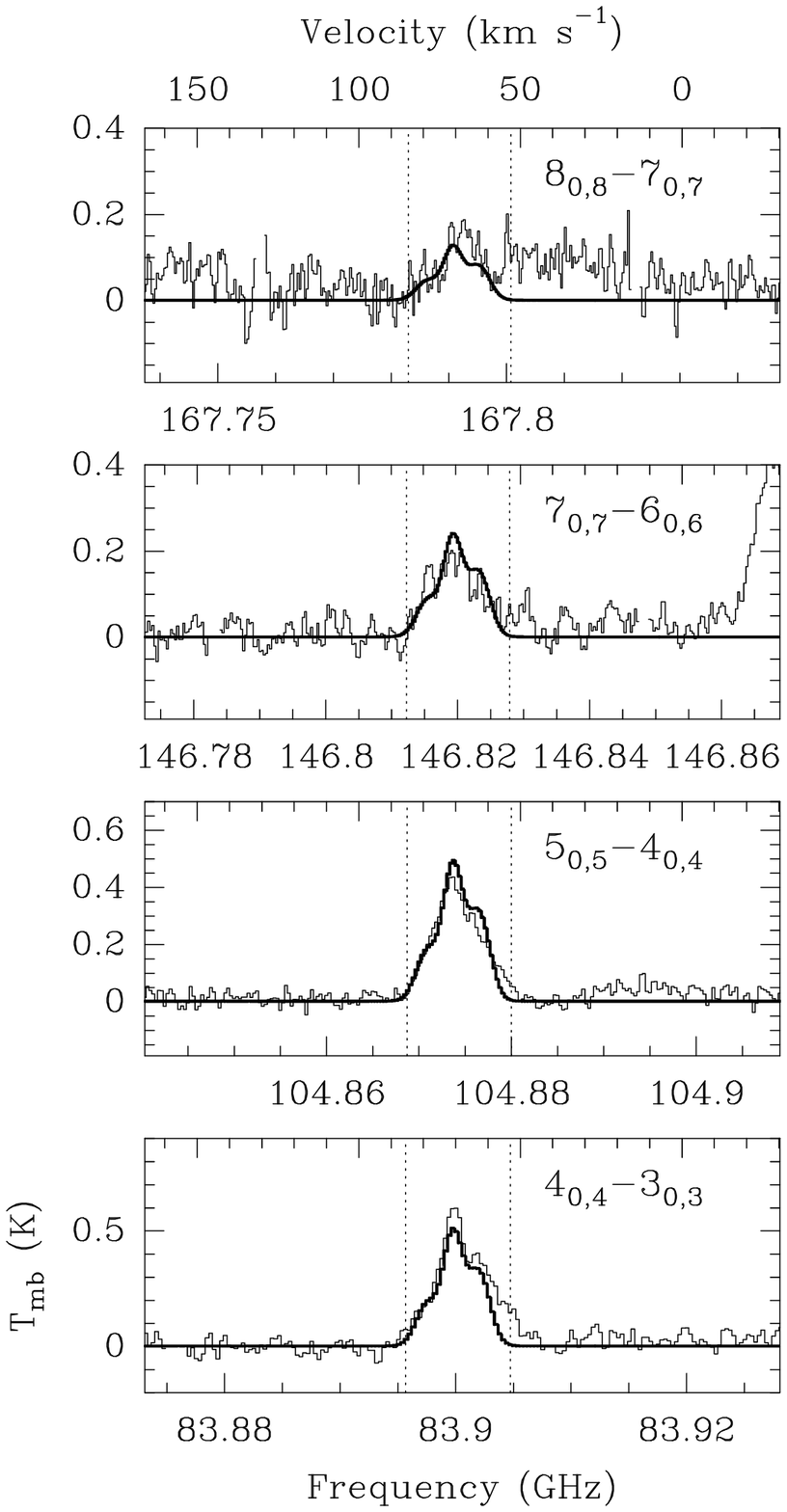}
\includegraphics[width=8.0cm]{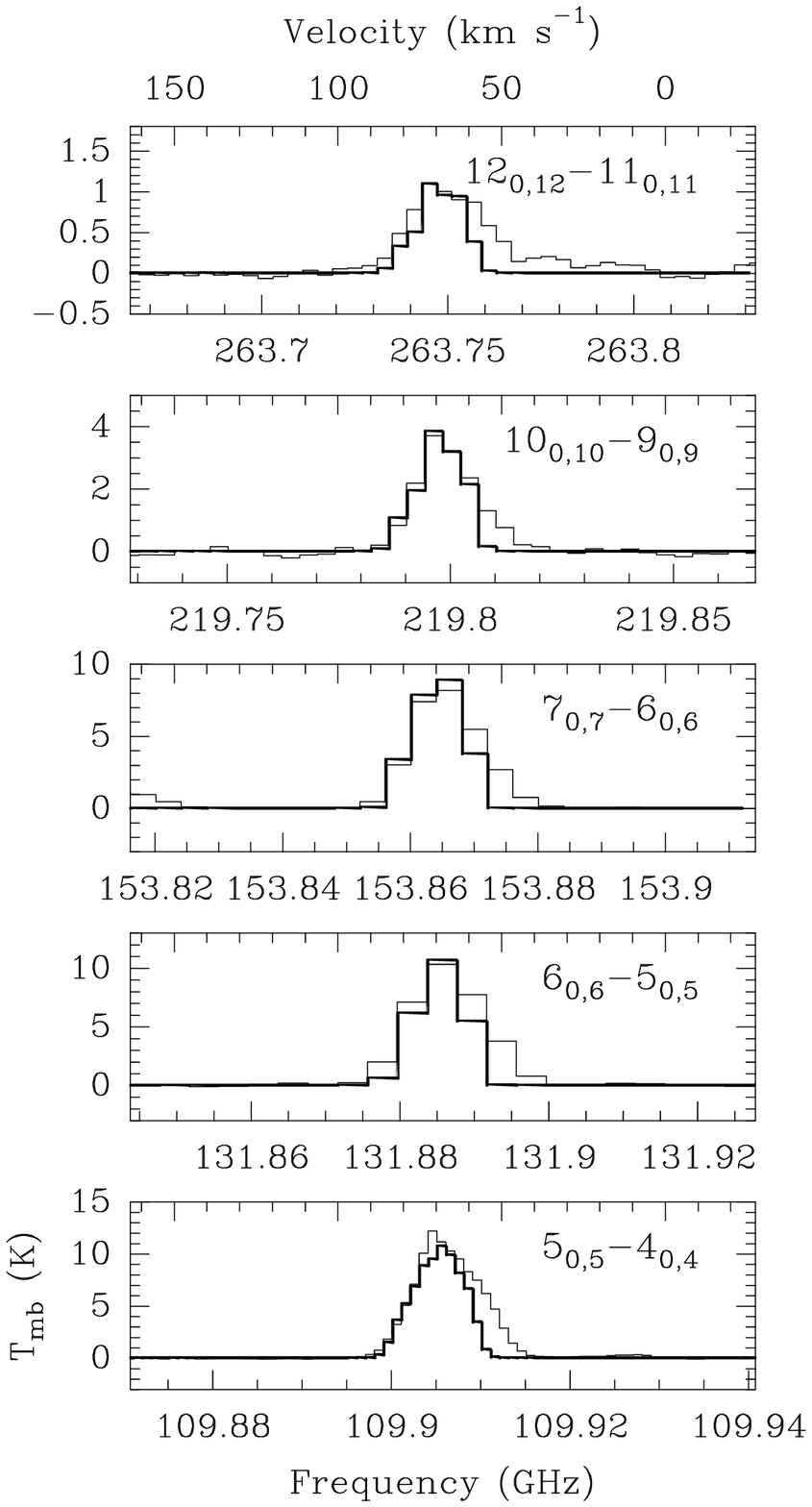}
\hspace{3cm} a) HOCN \hspace{7cm} b) HNCO
\caption{Detected lines of HOCN (this work) and HNCO (\cite{MRM2008}) observed with the IRAM 30m telescope toward Sgr B2(M)(20$^{\prime\prime}$,100$^{\prime\prime}$). In each panel, the observed spectrum (thin line) is shown in the main-beam brightness temperature scale, as well as the LTE synthetic spectrum of HOCN and HNCO, respectively (thick line). Three velocity components were needed to account for the observed line shape. The upper axis of each panel is a velocity axis, covering the same velocity range for all panels. The parameters of the LTE model are given in Table \ref{table_astropar}.}
\label{fig_M20,100}
\end{figure*}

\clearpage

\clearpage

\clearpage

\clearpage

\begin{figure*}
\includegraphics[width=8.0cm]{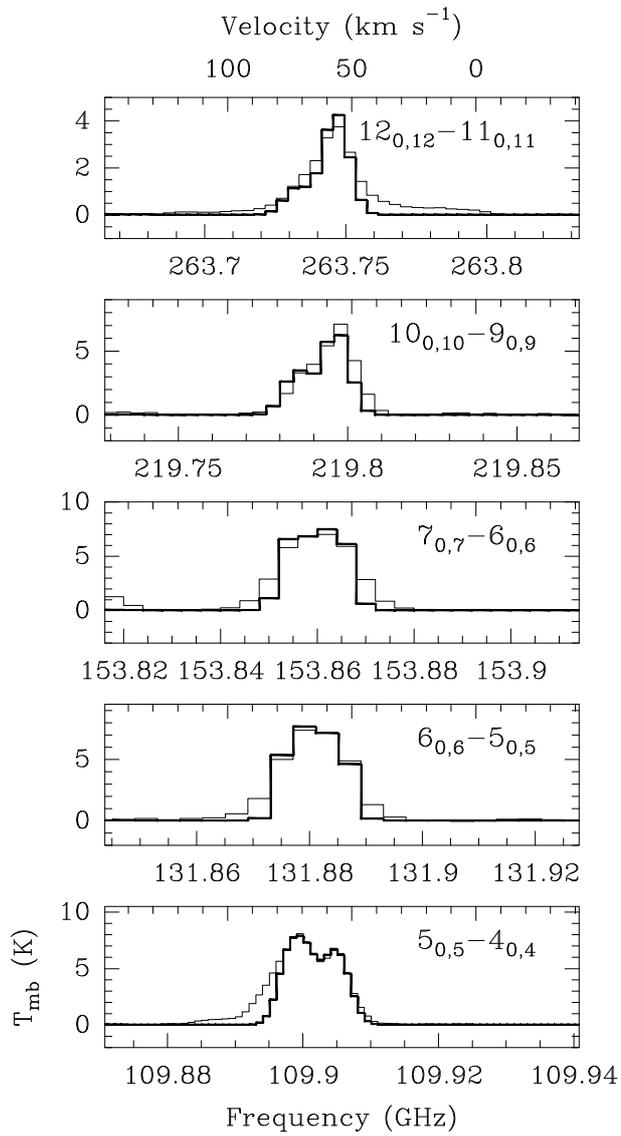}
\caption{Lines of HNCO observed toward Sgr B2(M)(-40$^{\prime\prime}$,0$^{\prime\prime}$) by \cite{MRM2008}. In each panel, the observed spectrum (thin line) is shown, as is the LTE synthetic spectrum of HNCO (thick line). The model was calculated by assuming the same two velocity components as for HOCN (see Fig. \ref{fig_offsetpositions} b) for comparison). The parameters of the LTE model are given in Table \ref{table_astropar}.}
\label{fig_HNCO_M-40,0}
\end{figure*}


\clearpage

\begin{table}
 \centering
 \caption{Coordinates of observed sources.}
\begin{tabular}{lccc}
\hline\hline
\multicolumn{1}{c}{Source} & \multicolumn{1}{c}{$\alpha_{J2000.0}$} & \multicolumn{1}{c}{$\delta_{J2000.0}$} & \multicolumn{1}{c}{$V_{LSR}$ (km/s)}\\  
\hline
Sgr B2(M) \hfill & $17^h47^m20.4^s$ & $-28^{\circ}23^{\prime}07^{\prime\prime}$ & $+62$\\
Sgr B2(N) \hfill & $17^h47^m20.0^s$ & $-28^{\circ}22^{\prime}19^{\prime\prime}$ & $+64$\\
Sgr B2(S)\hfill & $17^h47^m20.5^s$ & $-28^{\circ}23^{\prime}45^{\prime\prime}$ & $+61$\\
Sgr B2(M)(20$^{\prime\prime}$,100$^{\prime\prime}$) \hfill & $17^h47^m21.9^s$ & $-28^{\circ}21^{\prime}27^{\prime\prime}$ & $+68$\\
Sgr B2(M)(-40$^{\prime\prime}$,0$^{\prime\prime}$) \hfill & $17^h47^m17.4^s$ & $-28^{\circ}23^{\prime}07^{\prime\prime}$ & $+54$\\
Sgr B2(M)(20$^{\prime\prime}$,-180$^{\prime\prime}$) \hfill & $17^h47^m21.9^s$ & $-28^{\circ}26^{\prime}07^{\prime\prime}$ & $+50$\\
\hline
\end{tabular}
\label{table_positions}
\end{table}

\clearpage

\begin{table*}
 \centering
 \caption{
 Observed transitions of cyanic acid, HOCN, toward Sgr~B2.
}
 \label{table_detecthocn}
 \vspace*{0.0ex}
 \begin{tabular}{lcrcrrrl}
\hline\hline
 \multicolumn{1}{c}{Source$^a$} & \multicolumn{1}{c}{Transition} & \multicolumn{1}{c}{Frequency$^b$} & \multicolumn{1}{c}{$\delta f^c$} & \multicolumn{1}{c}{$\sigma^d$} & \multicolumn{1}{c}{I$_{\mathrm{obs}}^e$} & \multicolumn{1}{c}{I$_{\mathrm{HOCN}}^e$} & \multicolumn{1}{c}{Comments} \\ 
  & & \multicolumn{1}{c}{\scriptsize (MHz)} & \multicolumn{1}{c}{\scriptsize (MHz)} & \multicolumn{1}{c}{\scriptsize (mK)} & \multicolumn{1}{c}{\scriptsize (K~km$/$s)} & \multicolumn{1}{c}{\scriptsize (K~km$/$s)} & \\ 
 \multicolumn{1}{c}{(1)} & \multicolumn{1}{c}{(2)} & \multicolumn{1}{c}{(3)} & \multicolumn{1}{c}{(4)} & \multicolumn{1}{c}{(5)} & \multicolumn{1}{c}{(6)} & \multicolumn{1}{c}{(7)} & \multicolumn{1}{c}{(8)} \\ 
 \hline 
 (20'',100'') & 4$_{0,4}$-3$_{0,3}$ &   83900.572 & 0.31 &   25 &        10.19(15) &        7.89 & no blend
 \\ 
 & 5$_{0,5}$-4$_{0,4}$ &  104874.679 & 0.31 &   19 &         7.38(11) &        7.60 & no blend
 \\ 
 & 7$_{0,7}$-6$_{0,6}$ &  146820.687 & 0.31 &   29 &         3.59(13) &        3.66 & no blend
 \\ 
 & 8$_{0,8}$-7$_{0,7}$ &  167792.311 & 0.31 &   40 &         2.94(17) &        1.94 & weak detection
 \\ 
 Sgr B2(M) & 4$_{0,4}$-3$_{0,3}$ &   83900.572 & 0.31 &    8 &         2.13(05) &        1.67 & no blend
 \\ 
 & 5$_{0,5}$-4$_{0,4}$ &  104874.679 & 0.31 &   10 &         2.67(05) &        2.20 & partial blend with $^{13}$CH$_3$CH$_2$CN + U-line?
 \\ 
 & 7$_{0,7}$-6$_{0,6}$ &  146820.687 & 0.31 &   18 &         4.50(08) &        2.90 & partial blend with U-line?
 \\ 
 & 8$_{0,8}$-7$_{0,7}$ &  167792.311 & 0.31 &   43 &         6.56(17) &        3.65 & partial blend with U-line?
 \\ 
 Sgr B2(N) & 4$_{0,4}$-3$_{0,3}$ &   83900.572 & 0.31 &    9 &         4.66(05) &        3.74 & weak blend with CH$_3$CH$_3$CO
 \\ 
 & 5$_{0,5}$-4$_{0,4}$ &  104874.679 & 0.31 &   10 &        13.98(05) &        5.05 & strong blend with C$_2$H$_3$CN,v$_{11}$=1/v$_{15}$=1 and $^{13}$CH$_3$CH$_2$CN
 \\ 
 & 7$_{0,7}$-6$_{0,6}$ &  146820.687 & 0.31 &   17 &        12.01(08) &        6.89 & blend with U-lines
 \\ 
 & 8$_{0,8}$-7$_{0,7}$ &  167792.311 & 0.31 &   42 &        16.57(18) &        8.66 & blend with H$^{13}$CCCN,v$_6$=1 and U-line
 \\ 
 Sgr B2(S) & 7$_{0,7}$-6$_{0,6}$ &  146820.687 & 4.00 &   21 &         5.03(39) &        3.68 & no blend
 \\ 
 & 8$_{0,8}$-7$_{0,7}$ &  167792.311 & 4.00 &   20 &         0.70(32) &        1.60 & uncertain detection
 \\ 
 (20'',-180'') & 7$_{0,7}$-6$_{0,6}$ &  146820.687 & 4.00 &   29 &         1.76(53) &        1.93 & uncertain detection
 \\ 
 & 8$_{0,8}$-7$_{0,7}$ &  167792.311 & 4.00 &   24 &         0.03(38) &        1.51 & uncertain detection
 \\ 
  (-40'',0'') & 7$_{0,7}$-6$_{0,6}$ &  146820.687 & 4.00 &   26 &         4.76(61) &        4.55 & no blend
 \\ 
 & 8$_{0,8}$-7$_{0,7}$ &  167792.311 & 4.00 &   36 &         5.17(73) &        4.13 & uncertain detection
  \\ 
 \hline
 \end{tabular}
 \begin{list}{}{}
 \item[$(a)$]{The J2000 equatorial offsets are given with respect to Sgr~B2(M).}
 \item[$(b)$]{Laboratory rest frequencies were taken from \citep{BGM2009b}.}
 \item[$(c)$]{Channel width.}
 \item[$(d)$]{Measured rms noise level in T$_{\mathrm{mb}}$ scale.}
 \item[$(e)$]{Integrated intensity in T$_{\mathrm{mb}}$ scale for the observed spectrum (col. 6) and the model of cyanic acid (col. 7). The integration limits are marked in Figs. \ref{fig_M} - \ref{fig_offsetpositions} by vertical dotted lines. The uncertainty in col. 6 is given in parentheses in units of the last digit and is purely statistical.}
 \end{list}
 \end{table*}

\clearpage

\begin{table*}
 \centering
 \caption{
 Parameters of our best-fit LTE models of cyanic acid and isocyanic acid.
}
 \label{table_astropar}
 \vspace*{0.0ex}
 \begin{tabular}{lcrrcrrrrcrrr}
 \hline\hline
  & & \multicolumn{5}{c}{HOCN} & & \multicolumn{4}{c}{HNCO} & \\ 
 \multicolumn{1}{c}{Source$^{a}$} & \multicolumn{1}{c}{V$_{\mathrm{lsr}}$} & \multicolumn{1}{c}{Size$^{b}$} & \multicolumn{1}{c}{T$_{\mathrm{rot}}$} & \multicolumn{1}{c}{N$_{\mathrm{HOCN}}$$^{c}$} & \multicolumn{1}{c}{FWHM} & \multicolumn{1}{c}{V$_{\mathrm{off}}$$^{d}$} & &  \multicolumn{1}{c}{T$_{\mathrm{rot}}$} & \multicolumn{1}{c}{N$_{\mathrm{HNCO}}$$^{e}$} & \multicolumn{1}{c}{FWHM} & \multicolumn{1}{c}{V$_{\mathrm{off}}$$^{d}$} & \multicolumn{1}{c}{R$^{f}$} \\ 
  & \multicolumn{1}{c}{\scriptsize (km~s$^{-1}$)} & \multicolumn{1}{c}{\scriptsize ($''$)} & \multicolumn{1}{c}{\scriptsize (K)} & \multicolumn{1}{c}{\scriptsize (cm$^{-2}$)} & \multicolumn{1}{c}{\scriptsize (km~s$^{-1}$)} & \multicolumn{1}{c}{\scriptsize (km~s$^{-1}$)} & & \multicolumn{1}{c}{\scriptsize (K)} & \multicolumn{1}{c}{\scriptsize (cm$^{-2}$)} & \multicolumn{1}{c}{\scriptsize (km~s$^{-1}$)} & \multicolumn{1}{c}{\scriptsize (km~s$^{-1}$)} & \\ 
 \multicolumn{1}{c}{(1)} & \multicolumn{1}{c}{(2)} & \multicolumn{1}{c}{(3)} & \multicolumn{1}{c}{(4)} & \multicolumn{1}{c}{(5)} & \multicolumn{1}{c}{(6)} & \multicolumn{1}{c}{(7)} & & \multicolumn{1}{c}{(9)} & \multicolumn{1}{c}{(10)} & \multicolumn{1}{c}{(11)} & \multicolumn{1}{c}{(12)} & \multicolumn{1}{c}{(13)} \\ 
 \hline
  Sgr~B2(M)        & 62 & 60.0 &   $12^{+8}_{-3}$ &   $7.5^{+3.5}_{-1.5}  \times 10^{12}$ & 14.0 & 0.0  & &   $12^{+3}_{-2}$ & $  1.7^{+1.3}_{-0.7} \times 10^{15}$ & 14.0 & 1.0 &  $230^{+270}_{-140}$ \\  
                   &    & 3.0  &  $100^{+60}_{-60}$ & $  7.0^{+8.0}_{-4.0} \times 10^{14}$ & 12.0 & 0.0  & &  [100] & $  6.0 \times 10^{16}$ & 12.0 & 1.0 &   $90^{+110}_{-50}$ \\ 
  SgrB2~(N)        & 64 & 60.0 &   [14] & $  7.0^{+1.0}_{-2.0} \times 10^{12}$ & 9.0 & 0.0   & &   $14^{+6}_{-4}$ & $  1.7^{+1.3}_{-0.7} \times 10^{15}$ & 9.0 & -1.0 & $ 240^{+360}_{-115}$ \\  
                   &    & 60.0 &   [14] & $  9.0^{+2.0}_{-1.0} \times 10^{12}$ & 12.0 & 11.0 & &   $14^{+6}_{-4}$ & $  1.2^{+2.8}_{-0.2} \times 10^{15}$ & 12.0 & 14.0 &  $130^{+370}_{-40}$ \\  
                   &    & 2.4  &  [200] & $  3.0 \times 10^{15}$ & 7.0 & 0.0   & &  [200] & $  1.3 \times 10^{18}$ & 7.0 & -1.0 &  430 \\  
                   &    & 1.9  &  [200] & $  3.7 \times 10^{15}$ & 7.0 & 10.0  & &  [200] & $  8.0 \times 10^{17}$ & 7.0 & 9.0 &  220 \\  
  Sgr~B2(S)        & 61 & 60.0 &    $7^{+5}_{-5}$ & $  4.2^{+1.0}_{-3.0} \times 10^{13}$ & 17.0 & -2.0 & &   23 & $1.8 \times 10^{15}$$^g$ &  $\cdots$    &   $\cdots$   &  $43^{+107}_{-8}$ \\
 (20$^{\prime\prime}$,100$^{\prime\prime}$)  & 68 & 60.0 &    $9^{+4}_{-2}$ & $  5.0^{+2.0}_{-2.0} \times 10^{12}$ & 8.0 & 11.0  & &  $14^{+2}_{-1}$ & $  8.0^{+4.0}_{-3.0} \times 10^{14}$ & 8.0 & 11.0 &  $160^{+240}_{-90}$ \\  
                   &    & 60.0 &    $9^{+1}_{-1.5}$ & $  1.2^{+0.2}_{-0.2} \times 10^{13}$ & 7.0 & 3.0    & &  $14^{+2}_{-2}$ & $  2.0^{+1.0}_{-0.5} \times 10^{15}$ & 7.0 & 3.0 &  $170^{+130}_{-63}$ \\  
                   &    & 60.0 &    $9^{+4}_{-2}$ & $  9.0^{+1.0}_{-1.0} \times 10^{12}$ & 8.0 & -5.0   & &  $14^{+2}_{-2}$ & $  1.9^{+1.4}_{-0.5} \times 10^{15}$ & 8.0 & -5.0 &  $210^{+228}_{-70}$ \\  
 (-40$^{\prime\prime}$,0$^{\prime\prime}$)   & 54 & 60.0 &   [28] & $  9.5^{+3.5}_{-2.5} \times 10^{12}$ & 12.0 & 2.5   &  & $28^{+2}_{-4}$ & $  2.1^{+0.3}_{-0.3} \times 10^{15}$ & 12.0 & 2.5 &  $220^{+123}_{-20}$ \\  
                   &    & 60.0 &   [15] & $  7.0^{+2.0}_{-2.0} \times 10^{12}$ & 12.0 & 18.2 &  &  $15^{+3}_{-2}$ & $  2.7^{+1.0}_{-0.7} \times 10^{15}$ & 12.0 & 18.2 &  $350^{+8}_{-168}$ \\  
 (20$^{\prime\prime}$,-180$^{\prime\prime}$) & 50 & 60.0 &   [15] & $  7.0^{+2.0}_{-2.0} \times 10^{12}$ & 14.0 & 0.0  & &  15 & $  1.4 \times 10^{15}$$^g$ &  $\cdots$  &  $\cdots$  & $200^{+80}_{-44}$  \\
 \hline
 \end{tabular}
 \begin{list}{}{}
 \item[$(a)$]{The offsets are given with respect to Sgr~B2(M).}
 \item[$(b)$]{Source diameter (FWHM). The size of the compact sources was derived from the modeling of HNCO. The size of the extended sources was fixed to 60$''$.}
 \item[$(c)$]{Column density of cyanic acid.}
 \item[$(d)$]{Velocity offset with respect to the systemic velocity of the source given in col.~(2).}
 \item[$(e)$]{Column density of isocyanic acid.}
 \item[$(f)$]{Column density ratio [HNCO]/[HOCN].}
 \item[$(g)$]{Column density values in \cite{MRM2008} were multiplied by 1.4 to account for beam efficiencies and beam filling factors.} 
 \end{list}
 \end{table*}

\end{document}